\documentclass[onecollarge,natbib]{svjour2}
\bibpunct{[}{]}{;}{n}{}{,} 
\smartqed  
\usepackage{graphicx}
\journalname{}
\begin{document}

\title{No-Core Shell Model for Nuclear Systems with Strangeness}


\author{\hbox{D.~Gazda \and J.~Mare\v{s} \and P.~Navr\'{a}til \and R.~Roth \and R.~Wirth}}


\institute{D.~Gazda \at
              ECT$^*$, Villa Tambosi, I-38123 Villazzano (Trento), Italy\\
              Nuclear Physics Institute, 25068 \v{R}e\v{z}, Czech Republic \\
              \email{gazda@ujf.cas.cz}
           \and
           J.~Mare\v{s} \at
              Nuclear Physics Institute, 25068 \v{R}e\v{z}, Czech Republic
           \and
           P. Navr\'{a}til \at
              TRIUMF, 4004 Wesbrook Mall, Vancouver, BC V6T 2A3, Canada
           \and
           R.~Roth \at
              Institut f\"ur Kernphysik, Technische Universit\"at Darmstadt, 64289 Darmstadt, Germany
           \and
           R.~Wirth \at
              Institut f\"ur Kernphysik, Technische Universit\"at Darmstadt, 64289 Darmstadt, Germany
          }

\date{}

\maketitle

\begin{abstract}
We report on a novel {\it ab initio} approach for nuclear few- and many-body
systems with strangeness. Recently, we developed a relevant no-core shell model
(NCSM) technique \cite{Na09} which we successfully applied in first calculations
of lightest $\Lambda$ hypernuclei. The use of a translationally invariant finite
harmonic oscillator (HO) basis allows us to employ large model spaces, compared
to traditional shell model calculations, and use realistic nucleon--nucleon
($NN$) and nucleon--hyperon ($NY$) interactions (such as those derived from EFT
\cite{Po06}). We discuss formal aspects of the methodology, show first
demonstrative results for  ${}_{\Lambda}^3$H, ${}_{\Lambda}^4$H and
${}^4_\Lambda$He, and give outlook.

\keywords{hypernucleus \and ab initio calculations \and no-core shell model \and chiral EFT}
\end{abstract}

\section{Introduction}
\label{sec:introduction}
One of the major goals of contemporary theoretical hypernuclear physics is to
calculate the properties of hypernuclei starting from the bare interactions
between nucleons and hyperons. However, the $NY$ interaction is not well
understood due to the very limited $NY$ scattering database. Hypernuclei
therefore provide an important source of constraints on and serve as a testing ground
for the $NY$ interaction, as well as the $YY$ interaction where no scattering
data are available. Chiral perturbation theory seems to be a promising bridge
which allows us to connect QCD in the strangeness sector with low-energy
hypernuclear physics \cite{Po06,Ha13}. Moreover, lattice QCD simulations are
approaching a level where they can provide additional constraints on the
baryon--baryon interaction \cite{Be12}. In the near future, a significant part of the
research program in new facilities, like J-PARC in Japan and FAIR in Germany,
will be devoted to strangeness physics. The proposed experiments will, in
part, focus on accurate measurements of energy spectra of $S=-1$ and $S=-2$
hypernuclei \cite{jparc,panda}. To properly connect the characteristics of
hypernuclei with the underlying $NY$ and $YY$ interactions  reliable
calculations with realistic interactions are necessary. Nevertheless, only very
few \emph{ab initio} or \emph{exact} calculations of hypernuclei are available
for 3- and 4-body systems \cite{No02,Fi02,Ga07,Ne02,Hi02}. For $A\ge 4$ only a
handful of \emph{ab initio} approaches are applicable.

In Section \ref{sec:methodology}, the methodology used in our calculations is
briefly introduced, and in Section \ref{sec:results} we present first
demonstrative results. Brief summary with outlook is given in Section
\ref{sec:summary}.

\section{Methodology}
\label{sec:methodology}
Our approach is based on the no-core shell model technique \cite{Na09} which we
have extended to incorporate strangeness degrees of freedom. The no-core shell
model technique is one of the most powerful and universal \emph{ab initio}
methods in nuclear structure calculations. In the NCSM, the total Hamiltonian of
the system is diagonalized in a \emph{finite} $A$-body HO basis which is
truncated by a maximal HO excitation energy $N_{\rm max}\hbar\Omega$ with
respect to the unperturbed ground state of the $A$-body system. In the present
calculations we utilize the relative Jacobi-coordinate HO basis, which enable
us to perform calculations in larger model spaces compared to calculations in a conventional
Slater-determinant HO basis since the center-of-mass degrees of freedom are
explicitly removed and a basis coupled to a total angular momentum and isospin
is used . Unlike standard shell model calculations, in the NCSM all particles
are treated as active. This allows us to employ realistic $NN$ and $NY$
interactions. In this work we use the chiral N${^3}$LO $NN$ interaction
\cite{En03} and the chiral LO $NY$ interaction with cutoff $\Lambda=600$~MeV
\cite{Po06}. In the following calculations the potentials are used in a ``bare''
form without constructing any effective interaction tailored to a specific model
space. The results of calculations are thus variational with respect to the
model space truncation.

\section{Results and Discussion}
\label{sec:results}

\begin{figure}
\centering
\includegraphics[width=0.77\textwidth]{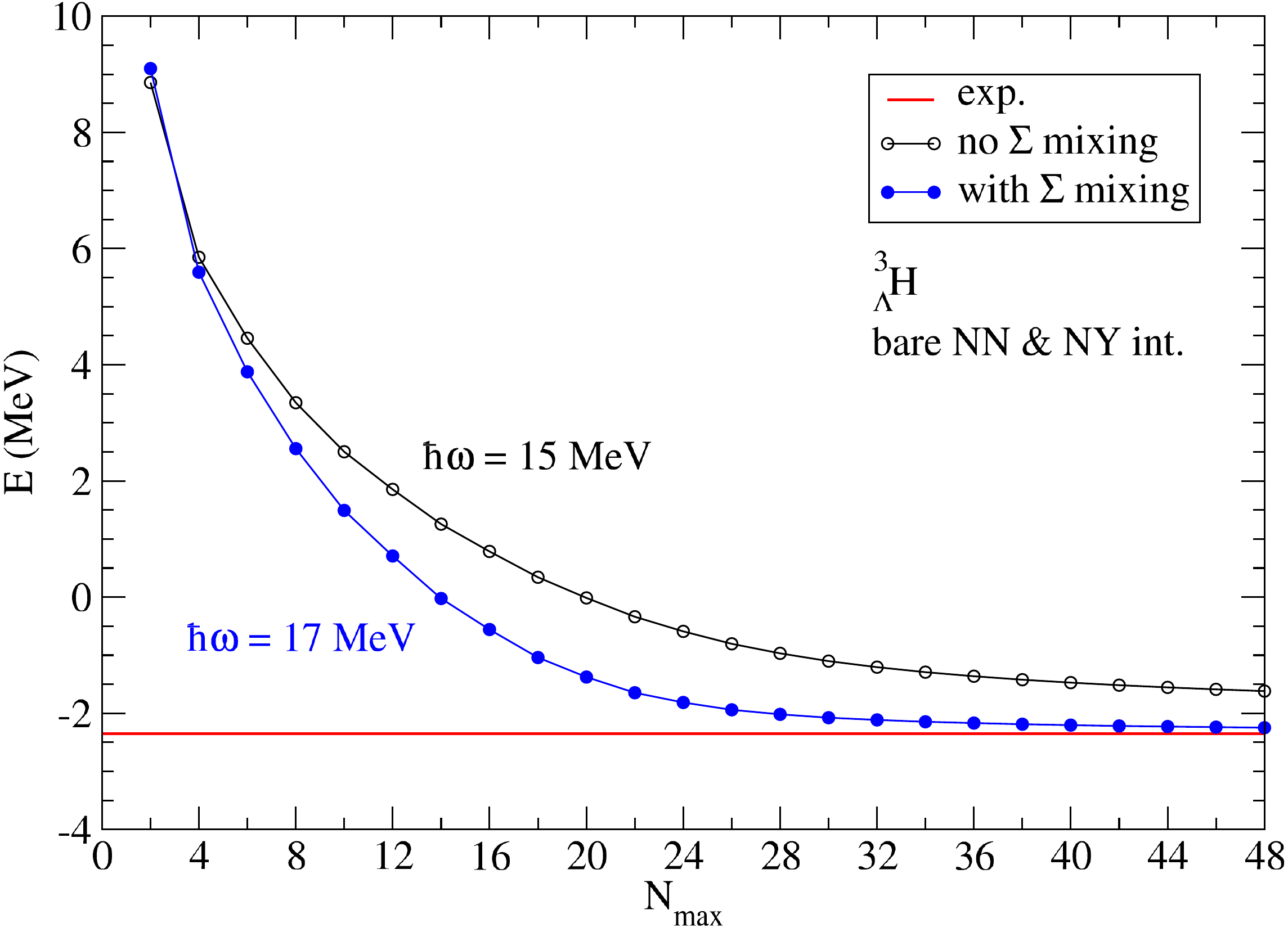}
\caption{(color online) The hypertriton ground state energy $E$ as a function of
the size of the model space $N_{\rm max}\hbar\omega$ calculated with the chiral
$NN$ and $NY$ forces. Results of calculations in the presence (absence) of the
explicit $\Lambda-\Sigma$ hyperon mixing in the hypernuclear state are denoted 
by full (open) circles for the optimal HO frequencies. The measured value of the
hypertriton ground state energy is shown by straight line.}
\label{fig:1}
\end{figure}

Our aim is to develop an \emph{ab initio} method for heavier $s$- and $p$-shell
hypernuclei. In our initial calculations we focus primarily on the
applicability of the NCSM to nuclear systems with nonzero strangeness,
particularly on the convergence properties and their dependence on the size of
the model space. We performed calculations of hypertriton ${}_\Lambda^3$H,
${}_\Lambda^4$H and ${}_\Lambda^4$He with realistic chiral $NN$ and $NY$
potentials.

In Fig.~\ref{fig:1}, the hypertriton ground state energy is shown as a function
of the size of the model space, calculated using the chiral $NN$ and $NY$
interactions. The results of calculations with the explicit $\Sigma$ hyperon
mixing in the hypertriton state (full circles) are compared to those calculated
in the absence of explicit $\Lambda - \Sigma$ mixing (open circles). The
inclusion of the $\Lambda - \Sigma$ hyperon mixing is essential in light
hypernuclear systems \cite{Ak00}. Due to the extremely weak binding of the
hypertriton, $E^{\rm (exp.)}({}^3_\Lambda{\rm
H})=-2.35\pm0.05$~MeV \cite{Da05}, the convergence with $N_{\rm max}$ is very
slow, as illustrated in the figure.
Preliminary extrapolation of the hypertriton ground state energy to the
infinite model space limit gives $E({}^3_\Lambda {\rm H})=-2.31 \pm
0.01$~MeV.

\begin{figure}
\centering
\includegraphics[width=0.77\textwidth]{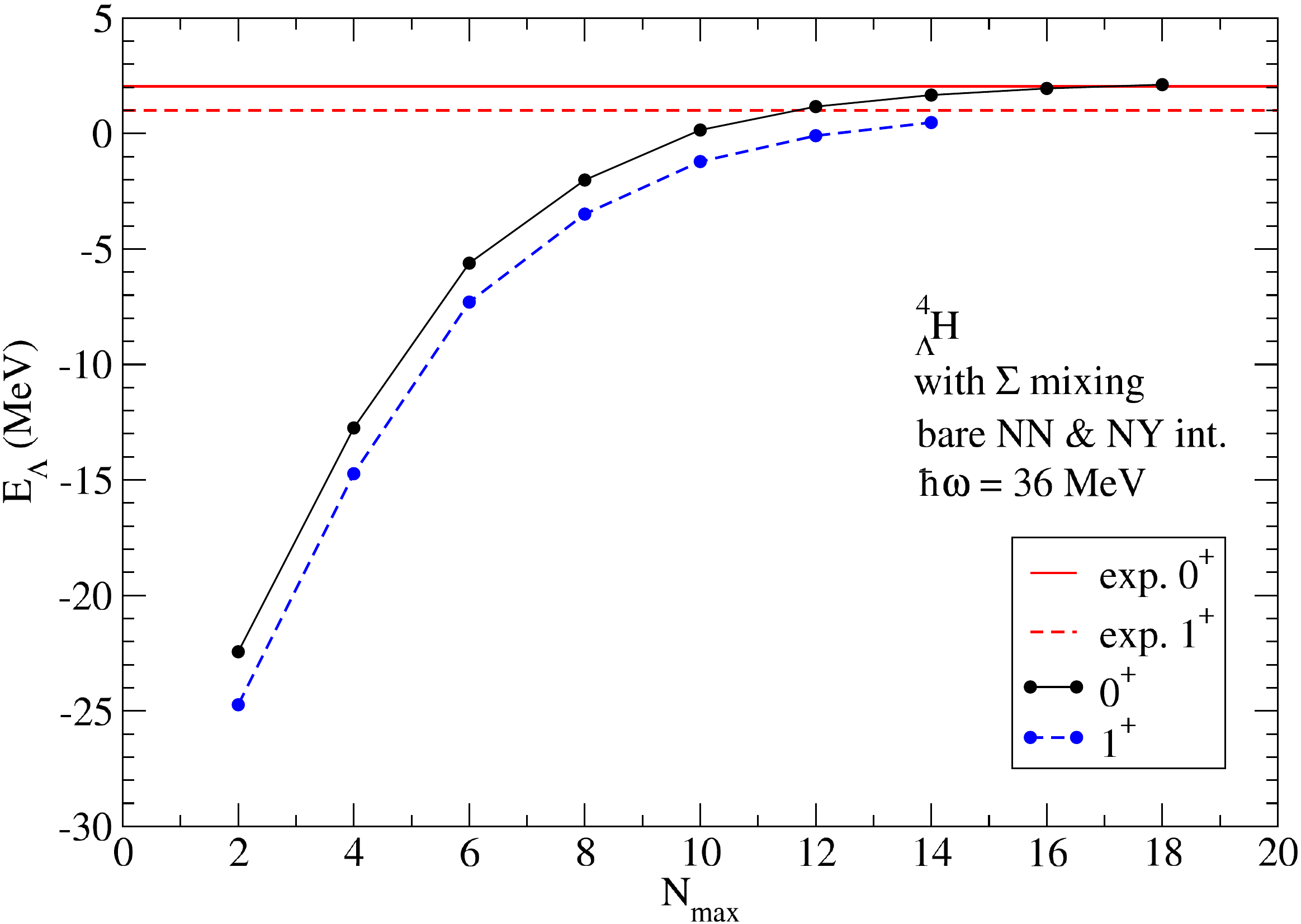}
\caption{(color online) The $\Lambda$ separation energy $E_\Lambda$ in the $0^+$
(full line) and $1^+$ (dashed line) states of ${}^4_\Lambda$H as a function of
the size of the model space $N_{\rm max}\hbar\omega$, for the optimal HO
frequency, calculated with the chiral $NN$ and $NY$ forces. The measured values
of the corresponding $\Lambda$ hyperon separation energies are shown by straight
lines.}
\label{fig:2}
\end{figure}

\begin{figure}
\centering
\includegraphics[width=0.77\textwidth]{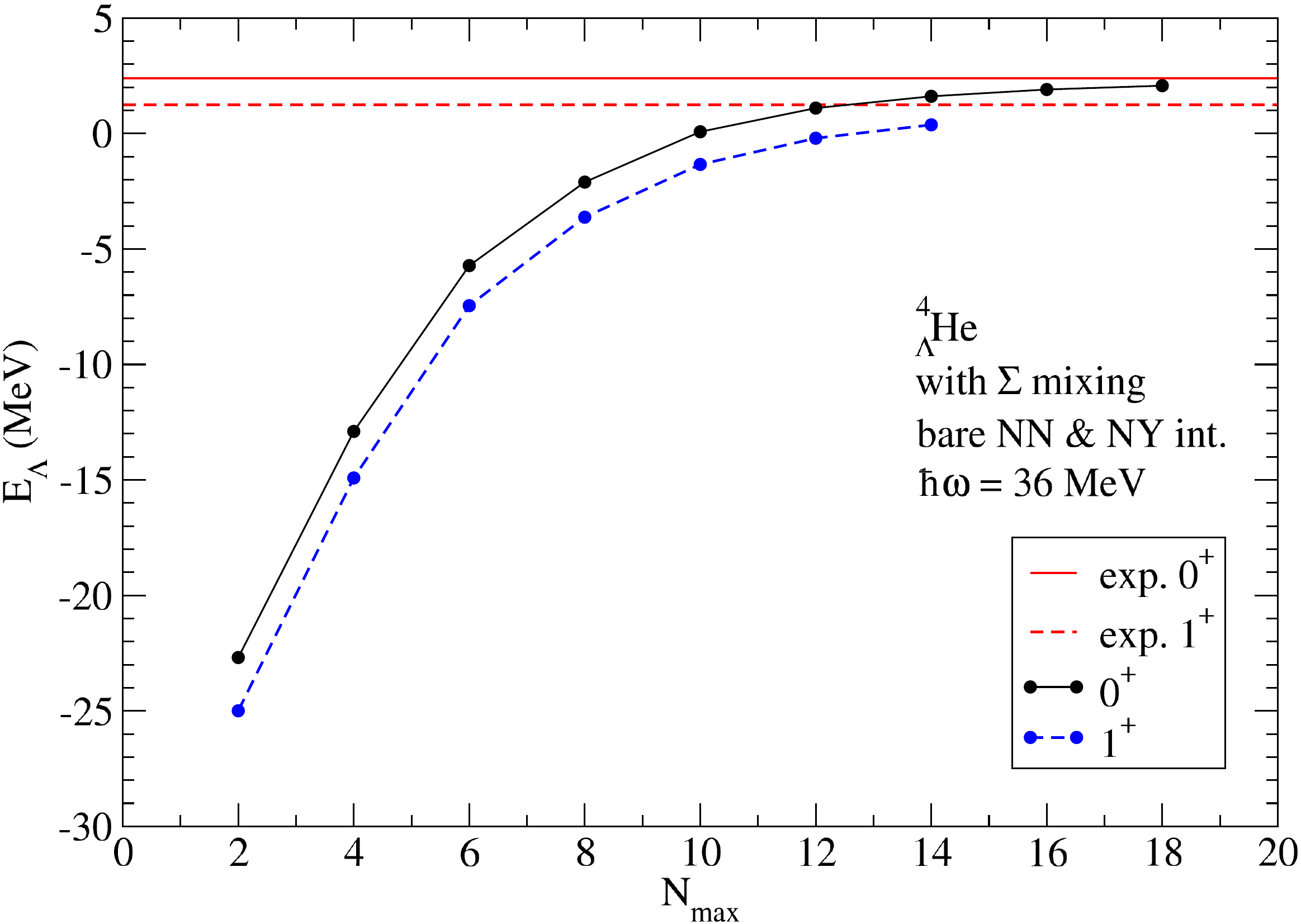}
\caption{(color online) The $\Lambda$ hyperon separation energy $E_\Lambda$ in
the $0^+$ (full line) and $1^+$ (dashed line) states of ${}^4_\Lambda$He as a
function of the size of the model space $N_{\rm max}\hbar\omega$, for the
optimal HO frequency, calculated with the chiral $NN$ and $NY$ forces. The
measured values of the corresponding $\Lambda$ hyperon separation energies are
shown by straight lines.}
\label{fig:3}
\end{figure}

Fortunately, in case of heavier and more bound and compact systems, such as
${}^4_\Lambda$H and ${}^4_\Lambda$He, the convergence with the size of the model
space is much faster, as shown in Figures \ref{fig:2} and \ref{fig:3}. Here the
$\Lambda$ hyperon separation energies, $E_\Lambda = E({}^{A-1}Z)-E({}^A_\Lambda
Z)$, in the $0^+$ and $1^+$ states of ${}^4_\Lambda$H and ${}^4_\Lambda$He are
shown as functions of the size of the model space, calculated with the chiral
$NN$ and $NY$ interactions. The results of calculations approach the measured
$\Lambda$ hyperon separation energies in the $0^+$($1^+$) state of
${}^4_\Lambda$H and ${}^4_\Lambda$He, $E_\Lambda^{\rm (exp.)}({}^4_\Lambda {\rm
H})=2.04\pm0.04\,(1.00\pm 0.06)$~MeV and $E_\Lambda^{\rm (exp.)}({}^5_\Lambda
{\rm He})=2.39\pm0.03\,(1.24\pm 0.05)$~MeV \cite{Da05}.

The results of our calculations demonstrate that the NCSM formalism presents a
powerful and promising tool to study nuclear systems with nonzero strangeness.
It is rather straightforward, and currently under development, to extend the
NCSM methodology to heavier, $A\ge 4$, hypernuclei. Nevertheless, more work is
needed to carry out systematic studies of the underlying interactions, perform
calculations in larger model spaces or employ effective interactions, and
include (possibly) $NNY$ forces.
Full details of the methodology and extrapolation of results to the
infinite model space limit will be discussed elsewhere.

\section{Summary}
\label{sec:summary}
The no-core shell model technique for nuclear systems with nonzero strangeness
was developed and applied in calculations of very light $\Lambda$ hypernuclei.
Our first results for ${}^3_\Lambda$H, ${}^4_\Lambda$H and ${}^4_\Lambda$He
using chiral $NN$ and $NY$ forces look promising and give reasonable description
of experimental data. The NCSM approach seems to be a viable candidate for the
\emph{ab initio} description of hypernuclei in the $A \ge 4$ mass region. In
case of ${}^3_\Lambda$H the convergence with the size of the model space is slow
as expected from the weak binding of the system. The convergence is much faster
for the heavier and more bound hypernuclei ${}^4_\Lambda$H and ${}^4_\Lambda$He.

In future we intend to extend our calculations to heavier $s$- and $p$-shell
hypernuclei and possibly study also other `exotic' systems, such as $\bar{K}$-
and $\eta$-nuclei.

\begin{acknowledgements}
We thank to Andreas Nogga for many useful discussions. This work was supported
in part by the GACR Grant No.\ 203/12/2126 and the NSERC Grant No.\ 401945-2011,
as well as by the EU initiative FP7, HadronPhysics3, under the SPHERE and
LEANNIS cooperation programs; and by HIC for FAIR and the DFG through SFB 634.
TRIUMF receives funding via a contribution through the Canadian National
Research Council.
\end{acknowledgements}

\end{document}